\documentclass[11pt]{article}

\usepackage{graphicx}
\usepackage{amssymb,amsmath}
\usepackage{cite}
\usepackage[symbol]{footmisc}
\usepackage{arydshln}
\usepackage{multirow}
\input{xy}\xyoption{all}

\numberwithin{equation}{section}

\newcommand{\pib}{\bar{\pi}}

\newcommand{\II}{\relax{\rm I\kern-.18em I}}

\newcommand{\rra}[1]{ \setbox1=\hbox{\kern10pt${#1}$\kern10pt}
  \,\vbox{\offinterlineskip\hbox to\wd1{\hfil\copy1\hfil}
    \kern 3pt\hbox to\wd1{\rightarrowfill}}\,}
\newcommand{\Bket}[1]{|\kern-.16em |#1\rangle\kern-.24em \rangle}
\newcommand{\Bovlap}[2]{\langle\kern-.24em \langle#1|\kern-.16em |#2\rangle\kern-.24em \rangle}

\title{\Large{\bf{Matrix Factorizations For Non-Abelian Orbifolds}}}
\author{Eleonora Dell'Aquila}

\date{}
\begin{document}
\ \vspace{0.3cm}\\
\begin{center}
{\Large\bf{Matrix Factorizations For Non-Abelian Orbifolds}}\vspace{1cm}\\
{\large Eleonora Dell'Aquila}
\end{center}
\vspace{0.1cm}
\begin{center}
{\it Perimeter Institute for Theoretical Physics \\
Waterloo, Ontario, N2L2Y5, Canada}\\
{\tt edellaquila@perimeterinstitute.ca}
\end{center}
\vspace{0.1cm}
\begin{abstract}
We discuss a prescription to construct fractional branes in Landau-Ginzburg orbifolds, with particular attention to the case of non-abelian orbifolds. We analyze in detail a $S_3$ orbifold and a $D_n$ orbifold and show how the computation of the spectrum can be carried out explicitly even when the factorizations involve higher rank matrices. 
\end{abstract}

\section{Introduction}
In recent years there has  been much progress in working with branes in Landau-Ginzburg models, in the framework of matrix factorizations. This approach was first proposed by M. Kontsevich and  was later introduced in the physics literature in \cite{K-Li,K-LiAG,Brunner}. 

In the simplest form, the idea is that the branes that preserve B-type supersymmetry can be classified by factorizations of the Landau-Ginzburg superpotential invoving matrices of arbitrary rank. Moreover, all correlators involving bulk and boundary chiral ring elements can be computed through a relatively simple procedure from the data of the factorization \cite{K-Li,Herbst}. The Landau-Ginzburg models that describe the tensor product of $\mathcal{N}=2$ minimal models have been studied with particular attention and all the boundary states known for minimal models have been mapped to matrix factorizations of  the corresponding Landau-Ginzburg potentials \cite{K-LiMM,Ashok,Brunner,Keller,BrunnerIII}. Moreover, the LG classification of B-branes has served as useful guidance in the study of minimal models: this has been the case, for example, for the construction of boundary states corresponding to permutation branes \cite{BGpermutation,permutationII,Fredenhagen, Caviezel}. 

B-branes in Landau-Ginzburg orbifolds were briefly discussed in \cite{K-LiMM} and were more systematically studied in \cite{Ashok}. Later, \cite{Walcher} filled in some gaps and proposed a formula for boundary correlators in orbifold models, generalizing the results of \cite{K-Li}. 
Essentially, if the LG model is orbifolded, in addition to a factorization of the superpotential one needs to specify a representation of the orbifold group.  The spectrum is then computed by solving some equivariance equations that implement the orbifold projection. 
However, although the formalism is quite general, the explicit examples considered in the physics literature have been mostly restricted to Landau-Ginzburg orbifolds that describe Gepner models. This class of examples is especially interesting, because not only is there vast availability of solid results on the CFT side, but these are also the models relevant for the Landau-Ginzburg/Calabi-Yau correspondence \cite{Ashok,AshokII,Aspinwall,Baumgartl,GLSMhori,Horiobs, Govindarajan} and several examples of mirror symmetry \cite{BrunnerC,Cornelius,io,K3LG,Knapp,Omer,KnSh}.  

The main purpose of this paper is to analyze explicitly a few examples of Landau-Ginzburg orbifolds that do not correspond to Gepner models. Mostly, we will consider non-abelian orbifolds, single out some class of factorizations - and representations of the orbifold group - and compute the spectrum of boundary operators for these factorizations. The general technology to do this is already in place, as mentioned above. The only new element here is perhaps the prescription to construct fractional branes. This is already implicit in the earlier work on $\mathbb{Z}_n$ orbifolds, but when considering non-abelian orbifold groups is becomes necessary to give a more systematic prescription to choose the representations associated to a given matrix factorization.  We will restrict our analysis to representations of the form 
\begin{equation*}
R^{(i)}\equiv r_i\otimes R^{(0)}\,,
\end{equation*}
where $r_i$ denotes the irreducible representations of the orbifold group and $R^{(0)}$ is a (generally reducible) representation that depends on the choice of factorization. This choice proves to be useful computationally, as most computations can be reduced to equations for matrices of the same size as $r_i$ instead of $R^{(i)}\,$. This simplification is one of the main results of the paper.

We will not discuss here any applications or comparisons with other models, but there are a few directions that it might be worth exploring. Some that come to mind are: boundary states for non-abelian orbifolds of minimal models; mirror symmetry for non-abelian Landau-Ginburg models; branes in permutation orbifolds of Calabi-Yau compactifications\footnote{In this case, however, the orbifold group would have to be abelian, since only cyclic permutations of minimal models preserve supersymmetry \cite{Klemm,Fuchs}.}. We hope to investigate some of these applications of the present work in the future.

\subsection*{Organization}
In Section \ref{general} we review some general facts about matrix factorizations and set up the notation and conventions used in the rest of the paper. We focus in particular on Landau-Ginburg orbifolds, and discuss the construction of fractional branes in these models. We also review the tensor product construction of factorizations and boundary operators introduced in \cite{Ashok}, which is crucial in carrying out most of the computations in this paper. 
In Section \ref{s2ex} we analyze in detail the orbifold $(W=X_1^n+X_2^n)/S_2\,$. Although this is a very simple and abelian orbifold, it contains many elements that return in more complex cases and it is therefore a good warm-up example.
In Section \ref{s3ex} we consider the orbifold $(W=X_1^n+X_2^n+X_3^n)/S_3\,$. Here we focus on a specific factorization and construct an orbit of fractional branes. We also show how the computation of the spectrum can be conceptually simplified and reduced to a few simple equations. The explicit results in matrix form are collected in the Appendix, both for completeness and comparison with the simple form of the answer obtained in the main text.
In Section \ref{dnex} we study our last example, the orbifold $(W=X_1^n+X_2^n)/D_n\,$. We again pick a specific factorization and compute the spectrum of the corresponding fractional branes. Section \ref{theend} contains a brief summary and some final comments. 
\subsection*{Acknowledgments}
I would like to thank Sujay Ashok for very helpful discussions and encouragement.
This research was supported by Perimeter Institute for Theoretical Physics. Research at
Perimeter Institute is supported by the Government of Canada through Industry Canada and
by the Province of Ontario through the Ministry of Research \& Innovation.
 
\section{Conventions and general formulae for matrix factorizations}\label{general}

In what follows we take $W(X_1,...,X_n)$ to be a polynomial function of the $n$ variables $X_i\,$. We also require $W$ to have only isolated critical points.

\paragraph{Unorbifolded LG model}

The topological B-branes of a Landau-Ginzburg model are classified by the matrix factorizations of the superpotential $W(X_1,\ldots,X_n)\,$ \cite{K-LiAG, Brunner, K-Li}. A brane is characterized by a matrix
\begin{equation}\label{fact}
D=\left[\begin{array}{cc}0&F\\ G&0\end{array}\right]\quad \text{such that} \quad D^2=\left[\begin{array}{cc} F\cdot G &0\\ 0&G\cdot F \end{array}\right]=\left[\begin{array}{cc} W_{LG}\cdot\II_k &0\\ 0&W_{LG}\cdot\II_k \end{array}\right]\, ,
\end{equation}
where $F$ and $G$ are rank $k$ matrices whose entries are polynomial in the variables $X_i$. If the model is not orbifolded this data is sufficient to compute the boundary chiral ring and all the topological observables relative to the brane. 

We distinguish even and odd boundary operators, of the form
\begin{align}
\phi=\left[\begin{array}{cc}\phi_-&0\\ 0&\phi_+\end{array}\right]&&
\text{and}&&
\psi=\left[\begin{array}{cc}0&\psi^-\\ \psi^+&0\end{array}\right]
\end{align}
respectively, where again $\phi^\pm,\psi^\pm$ are rank $k$ matrices with polynomial entries.  The spectrum of topological boundary changing operators between two branes $D_1$ and $D_2$ is found by solving the following equations:
\begin{subequations}\label{cohom}
\begin{align}
F_1\phi_+-\phi_-F_2=0 &\Big/ \{\phi_-=F_1\,\psi_+'+ \psi_-'\, G_2\,,\  \phi_+=G_1\,\psi_-'+\psi_+'\, F_2 \}  \\
F_1\psi_++\psi_-G_2=0&\Big/\{\psi_-=F_1\,\phi_+'-\phi_-'\,F_2\,,\  \psi_+= G_1\,\phi_-'-\phi_+'\,G_2\}\,.
\end{align} 
\end{subequations}
These are cohomology equations that specify the boundary chiral ring up to the addition of trivial operators, as specified in the curly brackets. 

Some information about the spectrum of boundary operators can be conveniently collected in the intersection matrix $I\,$, with
\begin{equation}
I_{ij}\equiv(\text{$\#$ even operators} - \text{$\#$ odd operators) between branes labeled by $i$ and $j$}\ .
\end{equation}
This is the same as the boundary Witten index,
$
I\equiv \text{Tr} (-1)^F\ .
$

As an example, let us consider the superpotential $W(X)=X^n\,$. If we take a factorization with $F=X\,,G=X^{n-1}\,$, then the spectrum of boundary preserving operators (i.e. $F_1=F_2$ and $G_1=G_2$ in \eqref{cohom}) consists of only one odd and one even operator, which we can take to have the form
\begin{align}\label{coh1var}
\phi=\II_2&&
\text{and}&&
\psi=\left[\begin{array}{cc}0&1\\ -X^{n-2}&0\end{array}\right]\,.
\end{align}


\paragraph{Tensor product of matrix factorizations}

The equations \eqref{cohom} are especially easy to solve if the rank of the factorization is $k=1\,$, i.e. if the superpotential $W$ can be factorized by polynomials. As it was shown in \cite{Ashok}, it is possible to construct some interesting higher rank factorizations from these through a tensor product construction and in this case it is sufficient to solve the equations for rank one to find the full boundary chiral ring. The tensor product of matrix factorizations in LG models maps to the usual tensor product of boundary states in the corresponding minimal models.  

We first note that if $F$ and $G$ are polynomials the matrix $D$ in \eqref{fact} can be rewritten as 
\begin{equation}
D=F \pi + G \pib \,,
\end{equation}
with
\begin{align}
\pi=\left(\begin{matrix}0&1\\0&0\end{matrix}\right)\qquad\qquad
\pib=\left(\begin{matrix}0&0\\1&0\end{matrix}\right)\,.
\end{align}
The matrices $\pi$ and $\pib$ satisfy the relations
\begin{equation}
\pi^2=0\,,\quad\pib^2=0\,,\quad\{\pi,\pib\}=\II_2\,.
\end{equation}
We can also rewrite the boundary operators as
\begin{align}\label{piop}
\phi=\phi_-\,\pi\pib+\phi_+\,\pib\pi\,,\qquad\qquad\psi=\psi_-\,\pi+\psi_+\,\pib\,.
\end{align}

Now suppose we have a Landau-Ginzburg superpotential $W(X_1,X_2)\equiv W_1(X_1)+W_2(X_2)\,$. We can take two factorizations of $W_1$ and $W_2$ and construct a factorization of $W$ as follows. First we need to find some matrices that satisfy the relations
\begin{equation}\label{pialgebra}
\begin{split}
&\{\pi_1,\pi_2\}=0\,,\quad\{\pib_1,\pib_2\}=0\,,
\quad\{\pi_1,\pib_2\}=0\,\quad\{\pi_2,\pib_1\}=0\,,\\&\{\pi_1,\pib_1\}=\II_4\,,\quad\{\pi_2,\pib_2\}=\II_4\,.
\end{split}
\end{equation}
We can take 
\begin{align}\label{pis}
\pi_1=\left(\begin{matrix}0&0&1&0\\ 0&0&0&0\\0&0&0&0\\0&1&0&0\end{matrix}\right)\ \pi_2=\left(\begin{matrix}0&0&0&-1\\ 0&0&0&0\\0&1&0&0\\0&0&0&0\end{matrix}\right)\ \pib_1=\left(\begin{matrix}0&0&0&0\\ 0&0&0&1\\1&0&0&0\\0&0&0&0\end{matrix}\right)\ \pib_2=\left(\begin{matrix}0&0&0&0\\ 0&0&1&0\\0&0&0&0\\-1&0&0&0\end{matrix}\right)\,.
\end{align}
Then it is easy to check using the relations \eqref{pialgebra} that 
\begin{equation}\label{Dtens}
D=F_1\pi_1+F_2\pi_2+G_1\pib_1+G_2\pib_2
\end{equation}
squares to $W$ if $F_i G_i=W_i\,$.

The construction extends to the boundary operators. It is easier to see this in an example, so let us consider $W(X_1,X_2)\equiv W_1(X_1)+W_2(X_2)\equiv X_1^n+X_2^n\,$, which will be relevant later. Let us pick the factorizations $F_1=X_1\,,G_1=X_1^{n-1}$ of $W_1$ and $F_2=X_2\,,G_2=X_2^{n-1}$ of $W_2\,$.  When considered independently, each factorization has boundary operators of the form \eqref{coh1var}, which we can rewrite
\begin{align}
\phi=\pi\pib+\pib\pi=\II_2\,,\qquad\quad
\psi=\pi-X^{n-2}\,\pib\,,
\end{align}
as in \eqref{piop}. Then one can check that the boundary preserving operators of the tensor product factorization \eqref{Dtens} can be taken to be
\begin{equation}\label{optens}
\begin{split}
&\II_4=\pi_1\pib_1+\pib_1\pi_1=\pi_2\pib_2+\pib_2\pi_2\\
&\Psi_1=\pi_1-X_1^{n-2}\,\pib_1\,,\quad
\Psi_2=\pi_2-X_2^{n-2}\,\pib_2\,,\\
&\Phi\equiv\Psi_1\Psi_2\,,
\end{split}
\end{equation}
where now the $\pi$'s are the four by four matrices in \eqref{pis}.

It should be reasonably intuitive how to generalize the construction to more complicated examples. We refer the reader to \cite{Ashok} for more details.

\paragraph{Orbifolded LG model}

If the Landau-Ginzgurg model is is orbifolded, we also need to specify the action of the orbifold group $G$ on the brane. This is done by choosing a representation 
\begin{align}
&R(g)=\left[\begin{array}{cc}R_-(g)&0\\ 0&R_+(g)\end{array}\right] \ \label{Rdef}\\
\intertext{
such that}
&R(g)D(X_i)=D\big(\rho(g)(X_i)\big)R(g) \qquad \forall g \in G\ \,,\label{equiv}\\
\intertext{or}
&R_-(g)F(X_i)=F\big(\rho(g)(X_i)\big)R_+(g)\label{equivF}\\
&R_+(g)G(X_i)=G\big(\rho(g)(X_i)\big)R_-(g)\qquad \forall g \in G\ \label{equivG}\,,
\end{align}
where $\rho$ denotes the action of the orbifold group on the $X_i$'s. When it can be solved, this equivariance condition relates the representations $R_+$ and $R_-$ in \eqref{Rdef}, so that the we only have the freedom of choosing one $k$-dimensional representation (where $k$ is the rank of $F$ and $G$ in \eqref{fact}). 

If for a given factorization $D$ we cannot find a pair of representations $R_+$ and $R_-$ that solve the condition \eqref{equiv}, then we cannot  associate a B-brane to the factorization $D$ in the orbifold theory. It means that the brane associated with $D$ in the unorbifolded theory is not invariant under $G$ and therefore we need to sum over all images of $D$ to construct a brane in the orbifold. After doing that we can proceed as described above.

For two branes $(D_1,R_1)$ and $(D_2,R_2)$ the spectrum of boundary operators is still given by \eqref{cohom}, but in addition we need to impose the equivariance conditions
\begin{subequations}\label{equivmor}
\begin{align}
&R_{2\pm}(g)\phi_\pm(X_i)=\phi_\pm\big(\rho(g)(X_i)\big)R_{1\pm}(g)\label{equivb}\\
&R_{2\pm}(g)\psi_\pm(X_i)=\psi_\pm\big(\rho(g)(X_i)\big)R_{1\mp}(g)\qquad \forall g \in G\label{equivf}\ \,.
\end{align}
\end{subequations}
Note that  if \eqref{equiv} is satisfied, the conditions for $\phi_+$ and $\psi_+$ are redundant.


\paragraph{Fractional branes}

In an orbifold model it is of particular interest to consider the set of fractional branes, which are labeled by the irreducible representations of the orbifold group $G\,$. In this section we give a prescription to construct fractional branes in the framework of matrix factorizations. 

Any discrete group $G$ has at least one one-dimensional representation, the trivial representation,
so given a factorization $D$ we can look for at least one solution to the equivariance condition \eqref{equiv}.  Let us denote $R^{(0)}$ one such solution, with
\begin{equation}
R^{(0)}(g)=\left[\begin{matrix}R^{(0)}_-(g)&0\\0&R^{(0)}_+(g)\end{matrix}\right]\qquad \forall g \in G\,
\end{equation}
and $R^{(0)}_-$ related to $R^{(0)}_+$ through \eqref{equiv}. Both  $R^{(0)}_-$ and $R^{(0)}_+$ are $k$-dimensional representations of $G\,$, in general not irreducible, and their structure depends on $D\,$.

Once we have picked $R^{(0)}\,$, we can construct a list of representations $R^{(i)}$ labeled by the irreducible representations $r_i$ of $G$ as follows:
\begin{equation}\label{fracrep}
R^{(i)}_\pm(g)\equiv r_i(g)\otimes R^{(0)}_\pm(g)\,.
\end{equation}
If $r_i$ has dimension $d\,$, we need to consider a superposition of $d$ copies of the matrix factorization $D\,$. We introduce the notation
\begin{equation}
D+D\equiv \left[\begin{matrix}0&0&F&0\\0&0&0&F\\G&0&0&0\\0&G&0&0\end{matrix}\right]
\end{equation}
The fractional branes are then described by pairs
\begin{equation}
(\,\underbrace{D+\ldots +D}_{d_i}\,,\,R^{(i)}\,)\,,
\end{equation}
where $d_i$ is the dimension of the $i$-th irreducible representation of $G\,$.
 
Note that since 
\begin{equation}
\bigoplus_i R^{(i)}_\pm(g)\equiv \left(\bigoplus_i r_i(g)\right)\otimes R^{(0)}_\pm(g)\,,
\end{equation}
it is true that a regular brane can be obtained as a suitable superposition of fractional branes. This is a good check that the construction of fractional branes proposed above makes sense. 
We should also point out that although higher rank matrices are generally involved, the construction has the advantage of leading to a few simplifications that make the computation of the spectrum of boundary operators more tractable. We will see this in the examples presented in the following sections.

\section{$S_2$}\label{s2ex}

As a warm-up example, we will consider in some detail the $S_2$ orbifold
\begin{equation}
(W_{LG}=X_1^n+X_2^n)/(X_1\leftrightarrow X_2)\,.
\end{equation}
\paragraph{Polynomial factorizations}
The superpotential can be written as 
\begin{equation}
W=\prod_{k=1}^nY_1-\eta_k Y_2\,,
\end{equation}
with $\eta_k^n=-1\,$, so the simplest factorization we can consider is
\begin{equation}\label{lin}
F_\eta= X_1-\eta X_2\,, \qquad G_\eta=\prod_{\{\eta_k\}\setminus \eta} X_1-\eta_k X_2\,.
\end{equation}
There are $n$ such factorizations, for each choice of $\eta\,$, and it was shown in \cite{BGpermutation} that these are the most fundamental branes in the model, in the sense that all other B-branes can be constructed as bound states of these.

The group $S_2$ has only two irreducible representations, the trivial and the alternating representation, both one-dimensional, so in this example the fractional branes are associated with representations
\begin{equation}
R(\sigma)=\left[\begin{array}{cc}(-1)^{\alpha_-}&0\\ 0&(-1)^{\alpha_+}\end{array}\right]\,,\qquad \alpha_-,\alpha_+ \in \{0,1\} \ ,
\end{equation} 
where $\sigma$ is the element of $S_2$ that exchanges $X_1$ and $X_2\,$.
The condition \eqref{equiv} gives
\begin{equation}
(-1)^{\alpha_-} (X_1-\eta X_2)  = (-1)^{\alpha_+} (X_2-\eta X_1) \,.
\end{equation}
The only case in which it is possible to find a solution is when $n$ is odd and $\eta=-1\,$. For the factorization $F_{-1}$ we find two branes, labeled by  $\alpha_-=\alpha_+\in \{0,1\}$ according to \eqref{equiv}.
If we look at the spectrum of boundary operators for $F_{-1}$ we find that there are no odd operators compatible with this boundary condition. There are instead even operators that solve \eqref{cohom} with
\begin{equation}
\phi_-=\phi_+\in \frac{\mathbb{C}[X_1-X_2]}{(X_1-X_2)^{n-1}}\,.
\end{equation}
The equivariance conditions \eqref{equivb} show that the boundary operator with $\phi_-=\phi_+=(X_1-X_2)^l$ preserves the boundary representation $R_\alpha$ if $l$ is even and changes the representation to $R_{\alpha+1} $ if $l$ is odd.

\paragraph{Sum over images}The remaining factorizations $F_\eta$ in \eqref{lin} correspond to boundary conditions that are not invariant under $S_2\,$. However, we can consider a superposition of two branes in the same $S_2$ orbit, so that the result is invariant. We take 
\begin{equation}
F_{\eta,\eta^{-1}}=\left[\begin{matrix} X_1-\eta X_2&0\\0& X_1-\eta^{-1} X_2 \end{matrix}\right]\,.
\end{equation}
One solution to the equivariance equation \eqref{equiv} is given by
\begin{align}
R^{(0)}_{-}(\sigma)=\left[\begin{matrix} 0&1\\1&0 \end{matrix}\right]&&R^{(0)}_+(\sigma)=\left[\begin{matrix} 0&-\eta^{-1}\\-\eta&0 \end{matrix}\right]\,,
\end{align}
so following \eqref{fracrep} we can construct fractional branes corresponding to the factorization $F_{\eta,\eta^{-1}}$ by choosing the representations
\begin{align}
R_{-}(\sigma)=\left[\begin{matrix} 0&-(1)^\alpha\\(-1)^\alpha&0 \end{matrix}\right]&&R_+(\sigma)=\left[\begin{matrix} 0&(-1)^{\alpha+1}\eta^{-1}\\(-1)^{\alpha+1}\eta&0 \end{matrix}\right] \qquad \alpha \in\{0,1\}\,.
\end{align}

Let us consider the spectrum of boundary operators between these fractional branes. According to \eqref{cohom} there are no odd operators and the even ones are of the form 
\begin{equation}
\phi_-=\left[\begin{matrix} f_\eta &0\\0&f_{\eta^{-1}}\end{matrix}\right]\,,\qquad\qquad \phi_+=\left[\begin{matrix} g_\eta &0\\0&g_{\eta^{-1}}\end{matrix}\right]\,,
\end{equation}
with 
\begin{align}
&f_\eta=g_\eta\,,\quad f_\eta\in \mathbb{C}[X_1,X_2]\Big/\{F_\eta,\frac{W}{F_\eta}\} \\ 
&f_{\eta^{-1}} = g_{\eta^{-1}}\,,\quad f_{\eta^{-1}}\in \mathbb{C}[X_1,X_2]\Big/\{F_{\eta^{-1}},\frac{W}{F_{\eta^{-1}}}\}\,.
\end{align}
We have some freedom in choosing the form of $f_\eta$ and $f_{\eta^{-1}}$ up to the equivalence relation, but we also need to satisfy the equivariance condition, which in this case reads
\begin{equation}\label{etaequiv}
(-1)^{\alpha_2} f_{\eta^{-1}}(X_1,X_2) = (-1)^{\alpha_1} f_\eta(\sigma(X_1),\sigma(X_2))\,.
\end{equation} 
A set of solutions is
\begin{align}
&\alpha_1=\alpha_2&& f_\eta=f_{\eta^{-1}}=(X_1+X_2)^l\,,\quad l\leq n-2\\
&\alpha_1=\alpha_2+1 &&f_\eta=-f_{\eta^{-1}}=(X_1+X_2)^l\,,\quad l\leq n-2\,,
\end{align}
corresponding to boundary preserving and boundary changing operators respectively. One can check that the other solutions of \eqref{etaequiv} are equivalent to these, since $X_1-X_2 $ can be expressed in terms of $X_1+X_2$ and $F_\eta$ or $F_{\eta^{-1}}\,$.

\paragraph{Tensor product factorizations} So far we have considered factorization of the kind \eqref{lin}. The analysis goes similarly for the other rank  one factorizations. Let us turn instead to the tensor product factorizations. To respect the permutation symmetry - and solve \eqref{equiv} - we consider only the tensor product of factorizations of the same degree $l\,$, that is
\begin{equation}\label{S2fact}
D=X_1^l\pi_1+X_2^l \pi_2 +X_1^{n-l}\,\pib_1+X_2^{n-l}\,\pib_2\,
\end{equation}
as in  \eqref{Dtens}. Using \eqref{pis} we get 
\begin{equation}\label{S2factM}
F=\left( \begin{matrix}X_1^l&-X_2^l\\X_2^{n-l}&X_1^{n-l} \end{matrix}\right) \qquad G=\left( \begin{matrix}X_1^{n-l}&X_2^l\\-X_2^{n-l}&X_1^{l} \end{matrix}\right)\,.
\end{equation}

The equivariance condition \eqref{equiv} in this case simply requires that the matrix $R(\sigma)$ should exchange $\pi_1,\pib_1$ and $\pi_2,\pib_2$ in \eqref{S2fact}. A solution is 
\begin{equation}
R^{(0)}_-(\sigma)= \left(\begin{matrix}-1&0\,\\0&1\,\end{matrix}\right)\qquad R^{(0)}_+(\sigma)= \left(\begin{matrix}0&1\\1&0\end{matrix}\right)\,,
\end{equation}
so the fractional branes in this case are characterized by the representations
\begin{equation}\label{S2rep}
R_-(\sigma)= (-1)^\alpha\otimes R^{(0)}_-(\sigma)\qquad R_+(\sigma)=  (-1)^\alpha\otimes R^{(0)}_+(\sigma)\qquad \alpha\in\{0,1\}\,.
\end{equation}

Let us focus on $l=1$ and look at the boundary operators, which for the unorbifolded case were worked out earlier in \eqref{optens}. Keeping into account the equivariance conditions, we find that the boundary preserving ($\alpha_1=\alpha_2$) operators are the identity $\II_4$ and $\Psi_1+\Psi_2\,$, with
\begin{align}
&(\Psi_1+\Psi_2)_-=\left[\begin{matrix}1&1\\X_2^2&-X_1^2\end{matrix}\right]\,,
&&(\Psi_1+\Psi_2)_+=\left[\begin{matrix}-X_1^2&-1\\-X_2^2&1\end{matrix}\right]\,,
\intertext{
while the boundary changing ($\alpha_1=\alpha_2+1$) operators are $\Psi_1\Psi_2\,$, with}
&(\Psi_1\Psi_2)_-=\left[\begin{matrix}0&1\\-X_1^2X_2^2&0\end{matrix}\right]\,,
&&(\Psi_1\Psi_2)_+= \left[\begin{matrix}0&X_1^2\\-X_2^2&0\end{matrix}\right]
\intertext{and $\Psi_1-\Psi_2\,$, with}
&(\Psi_1-\Psi_2)_-=\left[\begin{matrix}1&1\\X_2^2&-X_1^2\end{matrix}\right]\,,
&&(\Psi_1-\Psi_2)_+=\left[\begin{matrix}-X_1^2&-1\\-X_2^2&1\end{matrix}\right]\,.
\end{align}

Note that some significant simplifications have occurred.
Although the equivariance condition involves two by two matrices, the structure of the representations in \eqref{S2rep} is such that only the one-dimensional irreducible representation really enters the equation, once we have determined the $S_2$ charge of the boundary operators. This is because the equivariance condition can be rewritten as 
\begin{equation}
\mathcal{O}(X_i)=(-1)^{\alpha_1-\alpha_2}\,\left(R^{(0)}(\sigma(X_i))\right)^{-1}\mathcal{O}(\sigma(X_i))\,R^{(0)}(\sigma) \,,
\end{equation}
where $\mathcal{O}$ is any boundary operator.
Since $R^{(0)}$ exchanges $\pi_1 \leftrightarrow \pi_2\,$, this simply means
\begin{equation}
\mathcal{O}=(-1)^{\alpha_1-\alpha_2}\,\mathcal{O}_{\Psi_1\leftrightarrow \Psi_2} \,,
\end{equation}
and then the results above for the spectrum follow immediately without any explicit computation.
This observation becomes extremely useful for more complicated orbifolds, as we will see in the next section.

\section{$S_3$}\label{s3ex}

We consider the model 
\begin{equation}
\Big(W_{LG}=X_1^n+X_2^n+X_3^n\Big)\Big/S_3 \,,
\end{equation}
where the orbifold group $S_3$ is generated by 
\begin{align}
\sigma_{(12)}: X_1\leftrightarrow X_2 \qquad\text{and} \qquad \sigma_{(23)} :  X_2\leftrightarrow X_3 \,.
\end{align}
For later use, let us note that any representation of $S_3$ needs to satisfy the relations
\begin{equation}\label{S3rel}
R_{(12)}^2=\II\,,\quad R_{(23)}^2=\II\,,\qquad (R_{(12)}R_{(23)})^3=\II\,.
\end{equation}
The irreducible representations are classified by Young diagrams as follows:
\begin{align}
&\raisebox{-1mm}{\begin{picture}(30,10)
\put(0,10){\line(1,0){30}}\put(0,0){\line(1,0){30}}\put(0,0){\line(0,1){10}}\put(10,0){\line(0,1){10}}\put(20,0){\line(0,1){10}}\put(30,0){\line(0,1){10}}
\end{picture}}\,:
&& r_{\kern-.10em\begin{picture}(9,3)
\put(0,3){\line(1,0){9}}\put(0,0){\line(1,0){9}}\put(0,0){\line(0,1){3}}\put(3,0){\line(0,1){3}}\put(6,0){\line(0,1){3}}\put(9,0){\line(0,1){3}}
\end{picture}}^{(12)}=+1
&&r_{\kern-.10em\begin{picture}(9,3)
\put(0,3){\line(1,0){9}}\put(0,0){\line(1,0){9}}\put(0,0){\line(0,1){3}}\put(3,0){\line(0,1){3}}\put(6,0){\line(0,1){3}}\put(9,0){\line(0,1){3}}
\end{picture}}^{(23)}=+1\\
&\raisebox{-2.5mm}{\begin{picture}(20,20)
\put(0,10){\line(1,0){20}}\put(0,0){\line(1,0){10}}\put(0,20){\line(1,0){20}}\put(0,0){\line(0,1){20}}\put(10,0){\line(0,1){20}}\put(20,10){\line(0,1){10}}
\end{picture}}\,: 
&&r_{\kern-.10em\begin{picture}(6,8)
\put(0,3){\line(1,0){6}}\put(0,0){\line(1,0){3}}\put(0,6){\line(1,0){6}}\put(0,0){\line(0,1){6}}\put(3,0){\line(0,1){6}}\put(6,3){\line(0,1){3}}
\end{picture}}^{(12)}=
\left(\begin{matrix}
1&0\\0&-1
\end{matrix}\right)
&&r_{\kern-.10em\begin{picture}(6,8)
\put(0,3){\line(1,0){6}}\put(0,0){\line(1,0){3}}\put(0,6){\line(1,0){6}}\put(0,0){\line(0,1){6}}\put(3,0){\line(0,1){6}}\put(6,3){\line(0,1){3}}
\end{picture}}^{(23)}=
-\frac{1}{2}\left(\begin{matrix}
1&\sqrt{3}\\\sqrt{3}&-1
\end{matrix}\right)
\\
&\raisebox{-4.5mm}{\begin{picture}(10,30)
\put(0,10){\line(1,0){10}}\put(0,0){\line(1,0){10}}\put(0,20){\line(1,0){10}}\put(0,30){\line(1,0){10}}\put(0,0){\line(0,1){30}}\put(10,0){\line(0,1){30}}
\end{picture}}\,: 
&&r_{\kern-.10em\begin{picture}(3,10.5)
\put(0,3){\line(1,0){3}}\put(0,0){\line(1,0){3}}\put(0,6){\line(1,0){3}}\put(0,9){\line(1,0){3}}\put(0,0){\line(0,1){9}}\put(3,0){\line(0,1){9}}
\end{picture}}^{(12)}=-1
&&r_{\kern-.10em\begin{picture}(3,10.5)
\put(0,3){\line(1,0){3}}\put(0,0){\line(1,0){3}}\put(0,6){\line(1,0){3}}\put(0,9){\line(1,0){3}}\put(0,0){\line(0,1){9}}\put(3,0){\line(0,1){9}}
\end{picture}}^{(23)}=-1
\end{align}

The simplest factorization to consider in this case is the tensor product of three factorizations of $X^n$ of the same degree. To avoid cluttering the equations, let us focus on linear factorizations:
\begin{equation}\label{DS3}
D=X_1\pi_1+X_2 \pi_2+X_3 \pi_3 +X_1^{n-1}\,\pib_1+X_2^{n-1}\,\pib_2+X_3^{n-1}\,\pib_3\,,
\end{equation}
where here the $\pi_i$'s and $ \pib_i$'s are eight by eight matrices that satisfy relations of the kind \eqref{pialgebra} - the explicit form of these matrices can be found in the Appendix. In matrix form, we get
\begin{equation}\label{S3fact}
F=\left[
\begin{matrix}
X_1&0&-X_2&X_3\\
0&X_1&-X_3^{n-1}&-X_2^{n-1}\\
\phantom{-}X_2^{n-1}&X_3&\phantom{-}X_1^{n-1}&0\\
-X_3^{n-1}&X_2&\phantom{-}0&\phantom{-}X_1^{n-1}
\end{matrix}
\right]\quad
G=\left[
\begin{matrix}
\phantom{-}X_1^{n-1}&0&X_2&-X_3\\
0&X_1^{n-1}&X_3^{n-1}&\phantom{-}X_2^{n-1}\\
-X_2^{n-1}&-X_3&X_1&\phantom{-}0\\
\phantom{-}X_3^{n-1}&-X_2&0&\phantom{-}X_1
\end{matrix}
\right]\ .
\end{equation}

Solving the equivariance condition \eqref{equiv} in this case amounts to finding two eight by eight matrices $R_{(12)}\equiv\Big[\begin{smallmatrix}R^{(0)}_-[(12)]&0\\0&R^{(0)}_+[(12)]\end{smallmatrix}\Bigr]$ and $R_{(23)}\equiv\Big[\begin{smallmatrix}R^{(0)}_-[(23)]&0\\0&R^{(0)}_+[(23)]\end{smallmatrix}\Bigr]$ that satisfy \eqref{S3rel} in addition to
\begin{equation}
\begin{split}
R_{(12)}\cdot \pi_1 = \pi_2 \cdot R_{(12)} \qquad \qquad R_{(12)}\cdot \pi_3 = \pi_3 \cdot R_{(12)} \\
R_{(23)}\cdot \pi_1 = \pi_1 \cdot R_{(23)} \qquad \qquad R_{(23)}\cdot \pi_2 = \pi_3 \cdot R_{(23)} 
\end{split}
\end{equation}
and similar relations for the matrices $\pib_i\,$.
A solution is given by
\begin{subequations}
\begin{align}
&R^{(0)}_-[(12)]=\left[
\begin{matrix}
1&0&0&0\\
0&0&0&-1\\
0&0&-1&0\\
0&-1&0&0
\end{matrix}
\right]&&
R^{(0)}_+[(12)]=\left[
\begin{matrix}
0&0&-1&0\\
0&-1&0&0\\
-1&0&0&0\\
0&0&0&1
\end{matrix}
\right]\\
&R^{(0)}_-[(23)]=\left[
\begin{matrix}
1&0&0&0\\
0&-1&0&0\\
0&0&0&-1\\
0&0&-1&0
\end{matrix}
\right]
&&R^{(0)}_+[(23)]=\left[
\begin{matrix}
1&0&0&0\\
0&-1&0&0\\
0&0&0&-1\\
0&0&-1&0
\end{matrix}
\right]\,.
\end{align}
\end{subequations}

Following \eqref{fracrep}, the fractional branes associated to the trivial and alternating representations can be constructed by choosing the representations
\begin{equation}\label{1d}
R[(12)]= (-1)^\alpha \otimes R^{(0)}[(12)] \qquad R[(23)]= (-1)^\alpha  \otimes R^{(0)}[(23)]\qquad \alpha\in \{0,1\}\,.
\end{equation}
In order to realize the two dimensional irreducible representation we need to take the superposition of two branes of the form \eqref{S3fact} and then choose the representation
\begin{equation}\label{2d}
R_\pm[(12)]=r^{(12)} _{\kern-.10em\begin{picture}(6,8)
\put(0,3){\line(1,0){6}}\put(0,0){\line(1,0){3}}\put(0,6){\line(1,0){6}}\put(0,0){\line(0,1){6}}\put(3,0){\line(0,1){6}}\put(6,3){\line(0,1){3}}
\end{picture}}\otimes R_\pm^{(0)}[(12)] \qquad R_\pm[(23)]= r^{(23)}_{\kern-.10em\begin{picture}(6,8)
\put(0,3){\line(1,0){6}}\put(0,0){\line(1,0){3}}\put(0,6){\line(1,0){6}}\put(0,0){\line(0,1){6}}\put(3,0){\line(0,1){6}}\put(6,3){\line(0,1){3}}\end{picture}}\otimes
R_\pm^{(0)}[(23)]\,.
\end{equation}
One can check that the matrices so obtained solve the equivariance condition for the factorization $D+D\,$.

Let us now compute the spectrum of boundary operators between these fractional branes. Throughout this section we will denote by $\mathcal{F}_\alpha$ the branes associated with the representations \eqref{1d}, and by $\mathcal{F}_{\kern-.10em\begin{picture}(6,8)
\put(0,3){\line(1,0){6}}\put(0,0){\line(1,0){3}}\put(0,6){\line(1,0){6}}\put(0,0){\line(0,1){6}}\put(3,0){\line(0,1){6}}\put(6,3){\line(0,1){3}}
\end{picture}}$ the brane associated with the representation \eqref{2d}. \

We start from the spectrum of the fractional branes $\mathcal{F}_\alpha\,$.  As in the $S_2$ example considered earlier, it is quite easy to determine the spectrum without having to solve the equivariance equation explicitly. We have the three operators  
\begin{equation}\label{psis}
\Psi_1=\pi_1-X_1^{n-2}\, \pib_1\,,\qquad\Psi_2=\pi_2-X_2^{n-2}\, \pib_2\,,\qquad\Psi_3=\pi_3-X_3^{n-2} \,\pib_3
\end{equation}
as building blocks and the equations \eqref{equiv} for a generic operator can be rewritten as
\begin{equation}
\begin{split}
&\mathcal{O}=(-1)^{\alpha_1-\alpha_2}\,\mathcal{O}_{\Psi_1\leftrightarrow \Psi_2}\\
&\mathcal{O}=(-1)^{\alpha_1-\alpha_2}\,\mathcal{O}_{\Psi_2\leftrightarrow \Psi_3}\,.
\end{split}
\end{equation}
Therefore, only completely symmetric or antisymmetric combinations of $\Psi$'s will be in the spectrum. In addition to the identity operator we get one boundary preserving operator from the linear combination $\Psi_1+\Psi_2+\Psi_3\,$, and two boundary changing operators from $\Psi_1\Psi_2+\Psi_2\Psi_3+\Psi_3\Psi_1$ and $\Psi_1\Psi_2\Psi_3\,$. We can of course check explicitly that these are the solutions of \eqref{equivmor}. The matrix expression of these operators and those listed below can be found in the Appendix. 

Next, we look at the fractional brane $\mathcal{F}_{\kern-.10em\begin{picture}(6,8)
\put(0,3){\line(1,0){6}}\put(0,0){\line(1,0){3}}\put(0,6){\line(1,0){6}}\put(0,0){\line(0,1){6}}\put(3,0){\line(0,1){6}}\put(6,3){\line(0,1){3}}
\end{picture}}\,$. The even operators that preserve this boundary condition have the form 
\begin{equation}
\phi\equiv \left( \begin{matrix} \phi^{(1)}_-&\phi^{(2)}_-&0&0\\ \phi^{(3)}_-&\phi^{(4)}_-&0&0\\0&0&\phi^{(1)}_+&\phi^{(2)}_+\\0&0&\phi^{(3)}_+&\phi^{(4)}_+\end{matrix}\right)\,,
\end{equation} 
where each entry is a four by four matrix and each operator $\phi^{(i)}\equiv \left(\begin{smallmatrix}\phi^{(i)}_-&0\\0&\phi^{(i)}_+\end{smallmatrix}\right)$ solves \eqref{cohom} with $D$ given in \eqref{DS3}. This means that each $\phi^{(i)}$ is either the identity operator $\II_8$ or an even combination of the operators $\Psi_1\,$, $\Psi_2$ and $\Psi_3$ in \eqref{psis}.
Similarly, the odd boundary operators have the form 
\begin{equation}
\psi\equiv \left(\begin{matrix}  0&0&\psi^{(1)}_-&\psi^{(2)}_-\\ 0&0&\psi^{(3)}_-&\psi^{(4)}_-\\\psi^{(1)}_+&\psi^{(2)}_+&0&0\\\psi^{(3)}_+&\psi^{(4)}_+&0&0\end{matrix}\right)\,,
\end{equation} 
where each $\psi^{(i)}\equiv \left(\begin{smallmatrix}0&\psi^{(i)}_-\\\psi^{(i)}_+&0\end{smallmatrix}\right)$ is equal to an odd combination of the operators $\Psi_i\,$.

With these conventions, we can write the equivariance conditions \eqref{equivb} and \eqref{equivf} as
\begin{subequations}\label{phieqs}
\begin{align}
\left(\begin{matrix}
1&0\\0&-1
\end{matrix}\right)\otimes R^{(0)}_\pm[(12)] \ \cdot \ \phi_\pm(X_i)\ &= \ \phi_\pm(\sigma_{(12)}(X_i)) \ \cdot \ \left(\begin{matrix}
1&0\\0&-1
\end{matrix}\right)\otimes R^{(0)}_\pm[(12)] \label{phi12}\\
\left(\begin{matrix}
1&\sqrt{3}\\\sqrt{3}&-1
\end{matrix}\right)\otimes  R^{(0)}_\pm[(23)] \ \cdot \ \phi_\pm(X_i) \ &= \ \phi_\pm(\sigma_{(23)}(X_i)) \ \cdot \ \left(\begin{matrix}
1&\sqrt{3}\\\sqrt{3}&-1
\end{matrix}\right)\otimes  R^{(0)}_\pm[(23)]\label{phi23}
\end{align}
\end{subequations}
and
 \begin{subequations}\label{psieqs}
 \begin{align}
\left(\begin{matrix}
1&0\\0&-1
\end{matrix}\right)\otimes R^{(0)}_\pm[(12)] \ \cdot \ \psi_\pm(X_i)\ &= \ \psi_\pm(\sigma_{(12)}(X_i)) \ \cdot \ \left(\begin{matrix}
1&0\\0&-1
\end{matrix}\right)\otimes R^{(0)}_\pm[(12)] \\
\left(\begin{matrix}
1&\sqrt{3}\\\sqrt{3}&-1
\end{matrix}\right)\otimes  R^{(0)}_\pm[(23)] \ \cdot \ \psi_\pm(X_i) \ &= \ \psi_\pm(\sigma_{(23)}(X_i)) \ \cdot \ \left(\begin{matrix}
1&\sqrt{3}\\\sqrt{3}&-1
\end{matrix}\right)\otimes  R^{(0)}_\mp[(23)]
\end{align}
\end{subequations}
 respectively.
 
 Let us look at the equations for $\phi$ first. It is useful to rewrite $\phi_\pm$ as
 \begin{equation}\label{trick}
 \left(\begin{matrix}
1&0\\0&0
\end{matrix}\right) \otimes \phi^{(1)}_\pm +  \left(\begin{matrix}
0&1\\0&0
\end{matrix}\right) \otimes \phi^{(2)}_\pm + \left(\begin{matrix}
0&0\\1&0
\end{matrix}\right) \otimes \phi^{(3)}_\pm + \left(\begin{matrix}
0&0\\0&1
\end{matrix}\right) \otimes \phi^{(4)}_\pm \,.
 \end{equation}
Then, using the property 
\begin{equation}
(A\otimes B) \cdot (C\otimes D) = (A\cdot C) \otimes (B \cdot D)
\end{equation}
of the matrix tensor product, the equations \eqref{phieqs} reduce to
\begin{multline*}
\left[\begin{matrix}
 \phantom{-}R^{(0)}_\pm[(12)] \cdot \phi^{(1)}_\pm(X_i) &  \phantom{-}R^{(0)}_\pm[(12)] \cdot \phi^{(2)}_\pm(X_i)\\
 - R^{(0)}_\pm[(12)]\cdot \phi^{(3)}_\pm(X_i) & -R^{(0)}_\pm[(12)] \cdot \phi^{(4)}_\pm(X_i)
\end{matrix}\right]=\nonumber\\
=\left[\begin{matrix}
\phi^{(1)}_\pm(\sigma_{(12)}(X_i))\cdot R^{(0)}_\pm[(12)] & - \phi^{(2)}_\pm(\sigma_{(12)}(X_i))\cdot R^{(0)}_\pm[(12)]\\
 \phi^{(3)}_\pm(\sigma_{(12)}(X_i))\cdot R^{(0)}_\pm[(12)] &-\phi^{(4)}_\pm(\sigma_{(12)}(X_i))\cdot R^{(0)}_\pm[(12)]
\end{matrix}\right]
\end{multline*}
and
\begin{multline*}
\left[\begin{matrix}
R^{(0)}_\pm[(23)] \cdot \big(\phi^{(1)}_\pm(X_i) + \sqrt{3}\, \phi^{(3)}_\pm(X_i)\big)&  R^{(0)}_\pm[(23)] \cdot \big(\phi^{(2)}_\pm(X_i)+ \sqrt{3}\,\phi^{(4)}_\pm(X_i)\big)\\
 R^{(0)}_\pm[(23)]\cdot \big(\sqrt{3}\,\phi^{(1)}_\pm(X_i)-\phi^{(3)}_\pm(X_i) \big)& R^{(0)}_\pm[(23)] \cdot \big(\sqrt{3}\,\phi^{(2)}_\pm(X_i)-\phi^{(4)}_\pm(X_i) \big)
\end{matrix}\right]=\nonumber\\
=\left[\begin{matrix}
\big(\phi^{(1)}_\pm(\sigma_{(23)}(X_i))+\sqrt{3}\,\phi^{(2)}_\pm(\sigma_{(23)}(X_i))\big)\cdot R^{(0)}_\pm[(23)] & \big(\sqrt{3}\,\phi^{(1)}_\pm(\sigma_{(23)}(X_i))-\phi^{(2)}_\pm(\sigma_{(23)}(X_i))\big)\cdot R^{(0)}_\pm[(23)]\\
\big(\phi^{(3)}_\pm(\sigma_{(23)}(X_i))+\sqrt{3}\,\phi^{(4)}_\pm(\sigma_{(23)}(X_i))\big)\cdot R^{(0)}_\pm[(23)] &\big(\sqrt{3}\,\phi^{(3)}_\pm(\sigma_{(23)}(X_i))-\phi^{(4)}_\pm(\sigma_{(23)}(X_i))\big)\cdot R^{(0)}_\pm[(23)]
\end{matrix}\right]\ .
\end{multline*}
This greatly simplifies if we use again the fact that $R^{(0)}_\pm[(12)] $ exchanges $\pi_1$ and $\pi_2$ and $R^{(0)}_\pm[(23)] $ exchanges $\pi_2$ and $\pi_3\,$. We get the conditions
\begin{align}\label{simpler}
&\left\{
\begin{aligned}
\phi^{(1)}&=\phantom{-\,}[\phi^{(1)}]_{\Psi_1\leftrightarrow\Psi_2}\\
\phi^{(2)}&=-\,[\phi^{(2)}]_{\Psi_1\leftrightarrow\Psi_2}\\
\phi^{(3)}&=-\,[\phi^{(3)}]_{\Psi_1\leftrightarrow\Psi_2}\\
\phi^{(4)}&=\phantom{-\,}[\phi^{(4)}]_{\Psi_1\leftrightarrow\Psi_2}
\end{aligned}\right.
&&\text{and}&&
\left\{
\begin{aligned}
\phi^{(1)}+\sqrt{3}\, \phi^{(3)}= [\phi^{(1)}]_{\Psi_2\leftrightarrow\Psi_3}+\sqrt{3}\,[\phi^{(2)}]_{\Psi_2\leftrightarrow\Psi_3}\\
\phi^{(2)}+\sqrt{3}\, \phi^{(4)}=\sqrt{3}\,[\phi^{(1)}]_{\Psi_2\leftrightarrow\Psi_3}-[\phi^{(2)}]_{\Psi_2\leftrightarrow\Psi_3}\\
\sqrt{3}\,\phi^{(1)}- \phi^{(3)}=[\phi^{(3)}]_{\Psi_2\leftrightarrow\Psi_3}+\sqrt{3}\,[\phi^{(4)}]_{\Psi_2\leftrightarrow\Psi_3}\\
\sqrt{3}\,\phi^{(2)}-\phi^{(4)}=\sqrt{3}\,[\phi^{(3)}]_{\Psi_2\leftrightarrow\Psi_3}-[\phi^{(4)}]_{\Psi_2\leftrightarrow\Psi_3}
\end{aligned}\right.
\end{align}
where $[\phi^{(1)}]_{\Psi_1\leftrightarrow\Psi_2}$ denotes the operator $\phi^{(1)}$ with $\Psi_1$ and $\Psi_2$ exchanged and so on.  These equations can be easily solved. The first set of equations requires the boundary operators to be of the form
\begin{equation}
\begin{aligned}
&\phi^{(1)}=a\, \II + b\, (\Psi_1\Psi_3+\Psi_2\Psi_3)&&\quad
\phi^{(4)}=a'\, \II + b'\, (\Psi_1\Psi_3+\Psi_2\Psi_3)\\
&\phi^{(2)}=c\, \Psi_1\Psi_2 + d\, (\Psi_1\Psi_3-\Psi_2\Psi_3)&&\quad
\phi^{(3)}=c'\, \Psi_1\Psi_2 + d'\, (\Psi_1\Psi_3-\Psi_2\Psi_3)
\end{aligned}
\end{equation}
and substituting this in the second set of equations we find a basis of solutions
\begin{subequations}\label{sol1}
\begin{align}
\text{i.\quad }& \phi^{(1)}=\phi^{(4)}=\II\,,\ \phi^{(2)}=\phi^{(3)}=\emptyset\nonumber\\
\text{ii.\quad }&\phi^{(1)}=\phi^{(4)}=\emptyset\,,\ \phi^{(3)}=-\phi^{(2)}=\Psi_1\Psi_2 +\Psi_2\Psi_3+\Psi_3\Psi_1\\
\text{iii.\quad } &\phi^{(1)}=-\phi^{(4)}=\sqrt{3}\,(\Psi_1\Psi_3+\Psi_2\Psi_3)\,, \ \phi^{(2)}=3\, \Psi_1\Psi_2 \,,\ \phi^{(3)}=\Psi_1\Psi_2 + 2\, (\Psi_1\Psi_3-\Psi_2\Psi_3)\nonumber\ .
\intertext{We can proceed analogously to solve the equations \eqref{psieqs} for the odd operators. We get exactly the same equations \eqref{simpler}, but this time for the operators $\psi^{(i)}\,$.  The solutions are}
\text{iv.\quad }&\psi^{(1)}=\psi^{(4)}=\Psi_1+\Psi_2+\Psi_3\,,\ \psi^{(2)}=\psi^{(3)}=\emptyset\nonumber\\
\text{v.\quad }&\psi^{(1)}=\psi^{(4)}=\emptyset\,,\ \psi^{(3)}=-\psi^{(2)}=\Psi_1\Psi_2\Psi_3\ .
\end{align}
\end{subequations}
The explicit matrix form of this operators can be found in the Appendix.

\begin{figure}
\begin{center}
$\xymatrix{
&&\mathcal{F}_{\kern-.10em\begin{picture}(6,8)
\put(0,3){\line(1,0){6}}\put(0,0){\line(1,0){3}}\put(0,6){\line(1,0){6}}\put(0,0){\line(0,1){6}}\put(3,0){\line(0,1){6}}\put(6,3){\line(0,1){3}}
\end{picture}}\ar@(dl,ur)[lldd]^{-1}\ar@(ul,ur)[]^{+3-2}\ar@(dr,ul)[ddrr]_{+1}&&\\\\
\mathcal{F}_0\ar@(ur,dl)@<1ex>[rruu]^{-1}\ar@(l,u)[]^{+1-1}\ar@<1ex>[rrrr]^{+1-1}&&&&\mathcal{F}_1\ar@(u,r)[]^{+1-1}\ar[llll]^{+1-1}\ar@(ul,dr)@<-1ex>[uull]_{+1}
}$\caption{Fractional branes in $(W=X_1^n+X_2^n+X_3^n)/S_3$ orbifold listing the number of even and odd boundary operators between them.}\label{quivS3}
\end{center}
\end{figure}
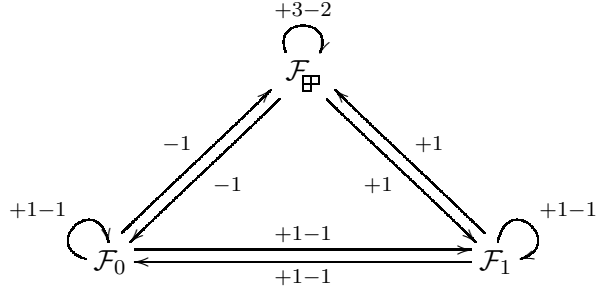

To complete the intersection matrix for the fractional branes, we need to compute the spectrum of operators between the branes $\mathcal{F}_\alpha$ and $\mathcal{F}_{\kern-.10em\begin{picture}(6,8)
\put(0,3){\line(1,0){6}}\put(0,0){\line(1,0){3}}\put(0,6){\line(1,0){6}}\put(0,0){\line(0,1){6}}\put(3,0){\line(0,1){6}}\put(6,3){\line(0,1){3}}
\end{picture}}$ and vice versa. The even and odd operators between $\mathcal{F}_\alpha$ and $\mathcal{F}_{\kern-.10em\begin{picture}(6,8)
\put(0,3){\line(1,0){6}}\put(0,0){\line(1,0){3}}\put(0,6){\line(1,0){6}}\put(0,0){\line(0,1){6}}\put(3,0){\line(0,1){6}}\put(6,3){\line(0,1){3}}
\end{picture}}$ have the form
\begin{equation}
\left( \begin{matrix} \phi^{(1)}_-&0\\ \phi^{(2)}_-&0\\0&\phi^{(1)}_+\\0&\phi^{(2)}_+\end{matrix}\right)\qquad\text{and}\qquad\left( \begin{matrix} 0&\psi^{(1)}_-\\ 0&\psi^{(2)}_-\\\psi^{(1)}_+&0\\\phi^{(2)}_+&0\end{matrix}\right)\,,
\end{equation} 
respectively. We find the equations
\begin{align}
&\left\{
\begin{aligned}
\phi^{(1)}&=(-1)^\alpha[\phi^{(1)}]_{\Psi_1\leftrightarrow\Psi_2}\\
\phi^{(2)}&=(-1)^{\alpha+1}\,[\phi^{(2)}]_{\Psi_1\leftrightarrow\Psi_2}\\
\end{aligned}\right.
&&\text{and}&&
\left\{
\begin{aligned}
\phi^{(1)}+\sqrt{3}\, \phi^{(2)}=-2\, (-1)^{\alpha}\,[\phi^{(1)}]_{\Psi_2\leftrightarrow\Psi_3}\\
\sqrt{3}\,\phi^{(1)}- \phi^{(2)}=-2 \,(-1)^{\alpha}\,[\phi^{(2)}]_{\Psi_2\leftrightarrow\Psi_3}\end{aligned}\right.
\end{align}
for the elements $\phi(i)$ and similarly for the $\psi^{(i)}\,$. We omit here most of the details because the computations are entirely analogous to the earlier case. 
We find that between $\mathcal{F}_0$ and $\mathcal{F}_{\kern-.10em\begin{picture}(6,8)
\put(0,3){\line(1,0){6}}\put(0,0){\line(1,0){3}}\put(0,6){\line(1,0){6}}\put(0,0){\line(0,1){6}}\put(3,0){\line(0,1){6}}\put(6,3){\line(0,1){3}}
\end{picture}}$ there are no even operators, while there is one odd operator of the form
\begin{subequations}\label{sol2}
\begin{align}
\text{i.\quad }&\psi^{(1)}=\Psi_1+\Psi_2-2\Psi_3\,,\quad\qquad\psi^{(2)}=\sqrt{3}\,(\Psi_2-\Psi_1)\,.
\intertext{Conversely, between $\mathcal{F}_1$ and $\mathcal{F}_{\kern-.10em\begin{picture}(6,8)
\put(0,3){\line(1,0){6}}\put(0,0){\line(1,0){3}}\put(0,6){\line(1,0){6}}\put(0,0){\line(0,1){6}}\put(3,0){\line(0,1){6}}\put(6,3){\line(0,1){3}}
\end{picture}}$ there are no odd operators and there is one even operator of the form}
\text{ii.\quad }&\phi^{(1)}=\Psi_1\Psi_3-\Psi_2\Psi_3+2\Psi_1\Psi_2\,,\quad\qquad\phi^{(2)}=\sqrt{3}\,(\Psi_1\Psi_3+\Psi_2\Psi_3)\,.
\end{align}
\end{subequations}
One can also check that these results hold also in the other direction, i.e. for the spectrum between $\mathcal{F}_{\kern-.10em\begin{picture}(6,8)
\put(0,3){\line(1,0){6}}\put(0,0){\line(1,0){3}}\put(0,6){\line(1,0){6}}\put(0,0){\line(0,1){6}}\put(3,0){\line(0,1){6}}\put(6,3){\line(0,1){3}}
\end{picture}}$ and  $\mathcal{F}_\alpha\,$, if we write the boundary operators in the form
\begin{equation}
\left( \begin{matrix} \phi^{(1)}_-&\phi^{(2)}_-&0 &0\\0&0&\phi^{(1)}_+&\phi^{(2)}_+\end{matrix}\right)\qquad\text{and}\qquad\left( \begin{matrix} 0&0&\psi^{(1)}_-&\psi^{(2)}_-\\\psi^{(1)}_+&\psi^{(2)}_+&0&0\end{matrix}\right)\,.
\end{equation} 

\ \\
To summarize, the intersection matrix is
\begin{equation}
I=\begin{pmatrix}+1-1&+1-1&-1\\+1-1&+1-1&+1\\-1&+1&+3-2\end{pmatrix}=\begin{pmatrix}0&0&-1\\0&0&1\\-1&1&1\end{pmatrix}\ .
\end{equation}

\section{$D_n$}\label{dnex}

We consider the orbifold
\begin{equation}
\Big(W=X_1^n+X_2^n\Big)\Big/D_n\,,\quad\qquad n>3\,,
\end{equation}
with the dihedral group $D_n$ is generated by $h\,,\,g_n$ acting as follows:
\begin{equation}
\begin{split}
h:\qquad &X_1\leftrightarrow X_2\\
g_n:\qquad &X_1 \to \omega_n X_1\\
&X_2\to \omega^{-1} X_2\,,\qquad \omega_n \equiv e^{2\pi i/n}\,.
\end{split}
\end{equation}
For simplicity, we will take $n$ to be even.
A presentation of the group through relations is
\begin{equation}\label{Drel}
R_h,R_g\  \big|\  R_h^2=1\,,\ R_g^n=1\,,\ R_hR_gR_h=R_g^{-1}\ .
\end{equation}
There are four one-dimensional irreducible representations labeled by $\alpha\,,\,\beta\in \{0,1\}\,$:
\begin{equation}
r^{(\alpha,\beta)}(g_n)=(-1)^\alpha \qquad \qquad r^{(\alpha,\beta)}(h)=(-1)^\beta\ ,
\end{equation}
and $\frac{n-2}{2}$ two-dimensional irreducible representations $r^{(\ell)}\,$, $\ell =1\,,\,\dots\,,\, \frac{n-2}{2}$ such that 
\begin{equation}
r^{(\ell)}(g_n)= \left(\begin{matrix}\omega_n^\ell&0\\0&\omega_n^{-\ell}\end{matrix}\right)\,,\qquad\quad r^{(\ell)}(h)=  \left(\begin{matrix}0&1\\1&0\end{matrix}\right)\ .
\end{equation}

We will consider the tensor product of two linear factorizations, i.e. a factorization of the form \eqref{S2factM} with $l=1\,$:
\begin{equation}\label{linfact}
F=\left( \begin{matrix}X_1&-X_2\\X_2^{n-1}&X_1^{n-1} \end{matrix}\right) \qquad G=\left( \begin{matrix}X_1^{n-1}&X_2\\-X_2^{n-1}&X_1 \end{matrix}\right)\,.
\end{equation}
We look for two matrices $R^{(0)}(h)$ and $R^{(0)}(g)$ that satisfy the equivariance condition \eqref{equiv} and the relations \eqref{Drel}. A solution is given by
\begin{subequations}
\begin{align}
&R^{(0)}_-(h)=\left[
\begin{matrix}
1&\phantom{-}0\\
0&-1
\end{matrix}
\right]&&
R^{(0)}_+(h)=\left[
\begin{matrix}
0&-1\\
-1&0
\end{matrix}
\right]\\
&R^{(0)}_-(g)=\left[
\begin{matrix}
1&0\\
0&1
\end{matrix}
\right]
&&R^{(0)}_+(g)=\left[
\begin{matrix}
\omega^{-1}&0\\
0&\omega
\end{matrix}
\right]\,.
\end{align}
\end{subequations}
Therefore the factorization \eqref{linfact} gives rise to four fractional branes $\mathcal{F}_{(\alpha,\beta)}$ with representations
\begin{equation}
R^{(\alpha,\beta)}(h)=(-1^\alpha)\otimes R^{(0)}(h)\qquad R^{(\alpha,\beta)}(g)=(-1^\beta)\otimes R^{(0)}(g)
\end{equation}
and $\frac{n-2}{n}$ fractional branes $\mathcal{F}_{(\ell)}$ with representations
\begin{equation}
\begin{split}
&R^{(\ell)}_-(h)=r^{(\ell)}(h)\otimes R^{(0)}_-(h)\qquad R^{(\ell)}_-(h)=r^{(\ell)}(h)\otimes R^{(0)}_+(h)\\
&R^{(\ell)}_-(g)=r^{(\ell)}(g)\otimes R^{(0)}_-(g)\qquad R^{(\ell)}(g)= r^{(\ell)}(g)\otimes R^{(0)}_+(g) \ .
\end{split}
\end{equation}

Let us start by computing the spectrum between the fractional branes $\mathcal{F}_{(\alpha,\beta)}\,$. The operators are made out of $\II_4\,,\,\Psi_1\,,\,\Psi_2$ as in \eqref{optens}, with the properties
\begin{subequations}\label{prop}
\begin{align}
\Psi_1(X_1)= R^{(0)}(h)\Psi_2(h(X_2)) R^{(0)}(h)&&\Psi_2(X_2)= R^{(0)}(h)\Psi_1(h(X_1)) R^{(0)}(h)\\
\omega^{-1} \Psi_1(X_1)= R^{(0)}(g)\Psi_1(g(X_1)) R^{(0)}(g)&&\omega \Psi_2(X_2)= R^{(0)}(h)\Psi_2(g(X_2)) R^{(0)}(h)
\end{align}
\end{subequations}
To solve the equivariance conditions \eqref{equivmor} we need to look for operators with charge $\pm$ under both $h$ and $g\,$. Such operators are $\II_4\,$, between $\mathcal{F}_{(\alpha,\beta)}$ and itself, and $\Psi_1\Psi_2\,$, between $\mathcal{F}_{(\alpha,\beta)}$ and $\mathcal{F}_{(\alpha+1,\beta)}\,$. There are no odd operators that preserve this class of boundary conditions.

We consider now the spectrum between two branes $\mathcal{F}_{(\ell')}$ and $\mathcal{F}_{(\ell)}\,$. 
Since these branes are constructed from the superposition of two factorizations \eqref{linfact}, the operators have the form 
\begin{equation}
\left( \begin{matrix} \phi^{(1)}_-&\phi^{(2)}_-&0&0\\ \phi^{(3)}_-&\phi^{(4)}_-&0&0\\0&0&\phi^{(1)}_+&\phi^{(2)}_+\\0&0&\phi^{(3)}_+&\phi^{(4)}_+\end{matrix}\right)\qquad \text{ans}\qquad 
 \left(\begin{matrix}  0&0&\psi^{(1)}_-&\psi^{(2)}_-\\ 0&0&\psi^{(3)}_-&\psi^{(4)}_-\\\psi^{(1)}_+&\psi^{(2)}_+&0&0\\\psi^{(3)}_+&\psi^{(4)}_+&0&0\end{matrix}\right)\,,
\end{equation} 
where  $\phi^{(i)}\equiv \left(\begin{smallmatrix}\phi^{(i)}_-&0\\0&\phi^{(i)}_+\end{smallmatrix}\right)$ and $\psi^{(i)}\equiv \left(\begin{smallmatrix}0&\psi^{(i)}_-\\\psi^{(i)}_+&0\end{smallmatrix}\right)$ are even and odd combinations respectively of $\Psi_1$ and $\Psi_2\,$. We use again the fact that these operators can be rewritten in the form \eqref{trick} and get the equations
\begin{subequations}
\begin{align}
&\mathcal{O}^{(1)}(X_i)= R^{(0)}(h)\mathcal{O}^{(4)}(h(X_i)) R^{(0)}(h)&&
\mathcal{O}^{(3)}(X_i)= R^{(0)}(h)\mathcal{O}^{(2)}(h(X_i)) R^{(0)}(h)\\
&\mathcal{O}^{(1)}(X_i)= \omega^{\ell'-\ell}R^{(0)}(g)\mathcal{O}^{(1)}(g(X_i)) R^{(0)}(g)&&
\mathcal{O}^{(4)}(X_i)= \omega^{\ell-\ell'}R^{(0)}(g)\mathcal{O}^{(4)}(g(X_i)) R^{(0)}(g)\\
&\mathcal{O}^{(2)}(X_i)= \omega^{-\ell'-\ell}R^{(0)}(g)\mathcal{O}^{(2)}(g(X_i)) R^{(0)}(g)&&
\mathcal{O}^{(3)}(X_i)= \omega^{\ell'+\ell}R^{(0)}(g)\mathcal{O}^{(3)}(g(X_i)) R^{(0)}(g)
\end{align}
\end{subequations}
where recall that $\ell,\ell'\in \{1\,,\ldots\,\frac{n-2}{2}\}\,$. 
Note that here $\mathcal{O}^{(i)}$ can stand for either $\phi^{(i)}$ or $\psi^{(i)}\,$, since we get the same equations in both cases. Making use of \eqref{prop} we can easily solve these equations and get the following spectrum:
\begin{subequations}
\begin{itemize}
\item[$\centerdot\ $] $\ell=\ell'\,$:\begin{equation}\begin{aligned}&\phi^{(1)}=\phi^{(4)}=\II_4\,,\ \phi^{(2)}=\phi^{(3)}=\emptyset\\&\phi^{(1)}=-\phi^{(4)}=\Psi_1\Psi_2\,,\ \phi^{(2)}=\phi^{(3)}=\emptyset\end{aligned}\end{equation}
\item[$\centerdot\ $] $\ell=\ell'+1\,$:\begin{equation}\psi^{(1)}=\Psi_1\,,\,\psi^{(4)}=\Psi_2\,,\ \psi^{(2)}=\psi^{(3)}=\emptyset\end{equation}\\
\item[$\centerdot\ $] $\ell=\ell'-1\,$:\begin{equation}\psi^{(1)}=\Psi_2\,,\,\psi^{(4)}=\Psi_1\,,\ \psi^{(2)}=\psi^{(3)}=\emptyset\end{equation}
\end{itemize}
\end{subequations}

Last, let us look at the spectrum between $\mathcal{F}_{(\alpha,\beta)}$ and $\mathcal{F}_{(\ell)}\,$. There are no even operators and for odd operators of the form
\begin{equation}
\psi\equiv\left( \begin{matrix} 0&\psi^{(1)}_-\\ 0&\psi^{(2)}_-\\\psi^{(1)}_+&0\\\phi^{(2)}_+&0\end{matrix}\right)\,,
\end{equation}
we find the equations
\begin{equation}
\begin{split}
&\psi^{(1)}(X_i)= (-1)^\alpha \,R^{(0)}(h)\psi^{(2)}(h(X_i)) R^{(0)}(h)\\
&\psi^{(1)}(X_i)= (-1)^\beta \omega^{-\ell} \,R^{(0)}(g)\psi^{(1)}(g(X_i)) R^{(0)}(g)\\
&\psi^{(2)}(X_i)= (-1)^\beta \omega^{\ell} \,R^{(0)}(g)\psi^{(2)}(g(X_i)) R^{(0)}(g)
\end{split}
\end{equation}
with solutions
\begin{subequations}
\begin{itemize}
\item[$\centerdot\ $]  $\alpha=0\,,\beta=0\,, \ell=1\,$: 
\begin{equation}
\psi^{(1)}=\Psi_2\,,\qquad\psi^{(2)}=\Psi_1
\end{equation}
\item[$\centerdot\ $]  $\alpha=1\,,\beta=0\,, \ell=1\,$: 
\begin{equation}
\psi^{(1)}=\Psi_2\,,\qquad\psi^{(2)}=-\Psi_1
\end{equation}
\item[$\centerdot\ $]  $\alpha=0\,,\beta=1\,, \ell=\frac{n}{2}-1\,$: 
\begin{equation}
\psi^{(1)}=\Psi_1\,,\qquad\psi^{(2)}=\Psi_2
\end{equation}
\item[$\centerdot\ $]  $\alpha=1\,,\beta=1\,, \ell=\frac{n}{2}-1\,$: 
\begin{equation}
\psi^{(1)}=\Psi_1\,,\qquad\psi^{(2)}=-\Psi_2
\end{equation}
\end{itemize}
\end{subequations}

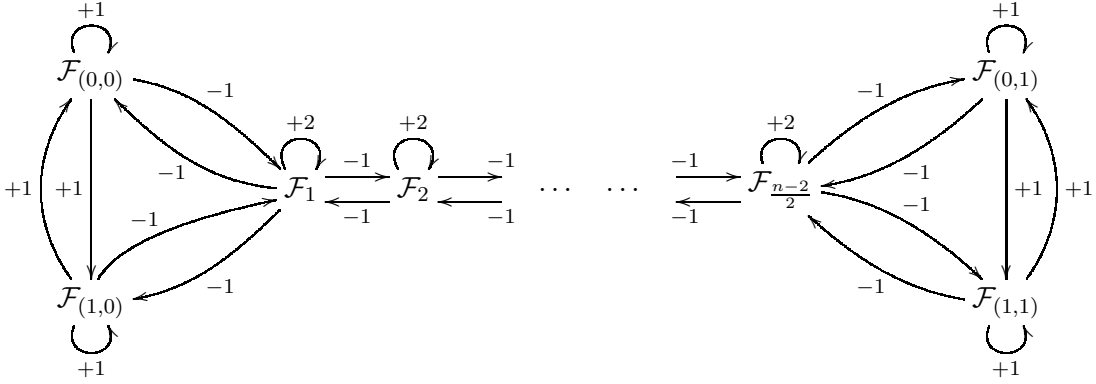
\begin{figure}
\begin{center}
$\xymatrix{
\mathcal{F}_{(0,0)}\ar@(r,ul)[rrd]^{-1}\ar[dd]_{+1}\ar@(ul,ur)[]^{+1}&&&&&&&\mathcal{F}_{(0,1)}\ar[dd]^{+1}\ar@(dl,r)[lld]^{-1}\ar@(ul,ur)[]^{+1}&\\&&\mathcal{F}_1\ar@(l,dr)[llu]^{-1}\ar@(dl,r)[lld]^{-1}\ar@(ul,ur)[]^{+2}\ar@<1ex>[r]^{-1}&\mathcal{F}_2\ar@<1ex>[l]^{-1}\ar@<1ex>[r]^{-1}\ar@(ul,ur)[]^{+2}&{\quad\ldots\quad\ldots\quad}\ar@<1ex>[l]^{-1}\ar@<1ex>[r]^{-1}&\mathcal{F}_{\frac{n-2}{2}}\ar@(r,ul)[rrd]^{-1}\ar@(ur,l)[urr]^{-1}\ar@(ul,ur)[]^{+2}\ar@<1ex>[l]^{-1}&\\\mathcal{F}_{(1,0)}\ar@(u,r)[rru]^(0.3){-1}\ar@(ul,dl)[uu]^{+1}\ar@(dl,dr)[]_{+1}&&&&&&&\mathcal{F}_{(1,1)}\ar@(l,dr)[llu]^{-1}\ar@(dl,dr)[]_{+1}\ar@(ur,dr)[uu]_{+1}
}$\caption{Fractional branes and boundary spectrum in $(W=X_1^n+X_2^n)/D_n$ orbifold}\label{quivDn}
\end{center}
\end{figure}

\noindent Similarly, in the other direction between $\mathcal{F}_{(\ell)}$ and  $\mathcal{F}_{(\alpha,\beta)}\,$, we find operators of the form 
\begin{equation*}
\psi\equiv\left( \begin{matrix} 0&0&\psi^{(1)}_-&\psi^{(2)}_-\\\psi^{(1)}_+&\psi^{(2)}_+&0&0\end{matrix}\right)\,
\end{equation*} 
with
\begin{subequations}
\begin{itemize}
\item[$\centerdot\ $] $\alpha=0\,,\beta=0\,, \ell=1\,$: 
\begin{equation}
\psi^{(1)}=\Psi_1\,,\qquad\psi^{(2)}=\Psi_2
\end{equation}
\item[$\centerdot\ $] $\alpha=1\,,\beta=0\,, \ell=1\,$: 
\begin{equation}
\psi^{(1)}=\Psi_1,,\qquad\psi^{(2)}=-\Psi_2
\end{equation}
\item[$\centerdot\ $] $\alpha=0\,,\beta=1\,, \ell=\frac{n}{2}-1\,$: 
\begin{equation}
\psi^{(1)}=\Psi_2\,,\qquad\psi^{(2)}=\Psi_1
\end{equation}
\item[$\centerdot\ $] $\alpha=1\,,\beta=1\,, \ell=\frac{n}{2}-1\,$: 
\begin{equation}
\psi^{(1)}=\Psi_2\,,\qquad\psi^{(2)}=-\Psi_1
\end{equation}
\end{itemize}
\end{subequations}
\ 

\noindent These results can be summarized in the intersection matrix as follows:\\ \\ 
\begin{equation*}
I=\begin{tabular}{ccccc:cccccccc}
&$(0,0)$&$(0,1)$&$(1,0)$&$(1,1)$&$\ell=1$&$\ldots$&$\ldots$&$\ldots$&$\ell=\frac{n-2}{2}$&\quad\qquad&\\\\
\multirow{9}{*}{
\begin{picture}(6,143)
\put(0,0){\line(1,0){6}}\put(0,0){\line(0,1){143}}\put(0,143){\line(1,0){6}}
\end{picture}
}
&$+1$&$0$&$+1$&$0$&$-1$&$0$&$\ldots$&$\ldots$&$0$&\multirow{9}{*}{
\begin{picture}(6,143)
\put(0,0){\line(1,0){6}}\put(6,0){\line(0,1){143}}\put(0,143){\line(1,0){6}}
\end{picture}
}
&&$(0,0)$\\
&$0$&$+1$&$0$&$+1$&$0$&$\dots$&$\ldots$&$0$&$-1$&&&$(0,1)$\\
&$+1$&$0$&$+1$&$0$&$-1$&$0$&$\ldots$&$\ldots$&$0$&&&$(1,0)$\\\vspace{0.1cm}
&$\phantom{(}0\phantom{)}$&$-1$&$0$&$+1$&$0$&$\ldots$&$\ldots$&$0$&$-1$&&&$(1,1)$\\\cdashline{2-10}\cdashline{13-13}
&$-1$&$\phantom{(}0\phantom{)}$&$-1$&$0$&$+2$&$-1$&$0$&$\ldots$&0&&&$\ell'=1$\\
&$0$&$\vdots$&$0$&$\vdots$&$-1$&$\ddots$&$\ddots$&$\ddots$&$0$&&&$\vdots$\\
&$\vdots$&$\vdots$&$\vdots$&$\vdots$&$0$&$\ddots$&$\ddots$&$\ddots$&$0$&&&$\vdots$\\
&$\vdots$&$0$&$\vdots$&$0$&$\vdots$&$\ddots$&$\ddots$&$\ddots$&$-1$&&&$\vdots$\\
&$0$&$-1$&$0$&$-1$&$0$&$0$&$0$&$-1$&$+2$&&&$\ell'=\frac{n-2}{2}$\\\\
\end{tabular}
\end{equation*}

\section{Conclusions}\label{theend}

In the last two sections we considered two examples of non-abelian Landau-Ginzburg orbifolds. We did not attempt to classify all B-branes in these models, but we picked a specific factorization of the superpotential and worked out the spectrum of the corresponding fractional branes. 

The methods used are not new, but there are a few facts to be learned from these examples. We have seen that there is a natural way to construct interesting fractional branes, associating representations of the form \eqref{fracrep} to the matrix factorizations. We have shown that for this choice of representation and for factorizations constructed as tensor products the computation of the spectrum can be reduced to the level of simplicity of a one-variable model. This can be quite significant for models of several variables, where the factorizations are in general of higher rank and lead to matrix cohomology equations that may appear intractable. Analogous observations for $\mathbb{Z}_n$ orbifolds have been crucial to analyze interesting LG models, such us the one describing the Gepner point in the moduli space of the quintc \cite{Ashok,AshokII,Horiobs}. We hope that the methods described here will be similarly useful. 

It would be interesting and reassuring to be able to make contact with known results from conformal field theory or geometry. Unfortunately there aren't as many results available as there are for abelian orbifolds and dualities between different models are much less clear. One possible route would be to try and reproduce the known results for ALE spaces \cite{OV,LercheALE,Johnson}, by considering a Landau-Ginzburg realization of a suitable toroidal compactification and probing the orbifold singularity with D0-branes. Another possibility would be to consider the free field representation of minimal models and construct boundary states in those variables to match the LG matrix factorizations. It could also be interesting to look at cyclic permutation orbifolds of minimal models in the moduli space of a Calabi-Yat compactification \cite{Klemm,Fuchs}.
We leave these investigations for future work.

\newpage

\begin{appendix}

\section{Some explicit results in matrix form from the $S_3$ example}
We collect here the explicit expressions of the operators that appear in Section \ref{s3ex}.
We take
\begin{align*}
\pi_1=\left(\begin{smallmatrix}&&&&\,1&\,0&\,0&\,0\\&&&&\,0&\,1&\,0&\,0\\&&&&\,0&\,0&\,0&\,0\\&&&&\,0&\,0&\,0&\,0\\\,0&\,0&\,0&\,0&&&&\\\,0&\,0&\,0&\,0&&&&\\\,0&\,0&\,1&\,0&&&&\\\,0&\,0&\,0&\,1&&&&\end{smallmatrix}\right)
&&\pi_2=\left(\begin{smallmatrix}&&&&0&0&-1&0\\&&&&0&0&\,0&0\\&&&&0&0&\,0&0\\&&&&0&1&\,0&0\\ 0&\,0&1&0&&&&\\0&\,0&0&0&&&&\\0&\,0&0&0&&&&\\0&-1&0&0&&&&\end{smallmatrix}\right)
&&&\pi_3=\left(\begin{smallmatrix}&&&&\,0&\,0&\,0&\,1\\&&&&\,0&\,0&\,0&\,0\\&&&&\,0&\,1&\,0&\,0\\&&&&\,0&\,0&\,0&\,0\\ 0&\,0&0&-1&&&&\\0&\,0&0&\,0&&&&\\0&-1&0&\,0&&&&\\0&\,0&0&\,0&&&&\end{smallmatrix}\right)\\
\pib_1=\left(\begin{smallmatrix}&&&&\,0&\,0&\,0&\,0\\&&&&\,0&\,0&\,0&\,0\\&&&&\,0&\,0&\,1&\,0\\&&&&\,0&\,0&\,0&\,1\\\,1&\,0&\,0&\,0&&&&\\\,0&\,1&\,0&\,0&&&&\\\,0&\,0&\,0&\,0&&&&\\\,0&\,0&\,0&\,0&&&&\end{smallmatrix}\right)
&&\pib_2=\left(\begin{smallmatrix}&&&&0&0&0&\,0\\&&&&0&0&0&-1\\&&&&1&0&0&\,0\\&&&&0&0&0&\,0\\ \,0&0&0&0&&&&\\\,0&0&0&1&&&&\\-1&0&0&0&&&&\\\,0&0&0&0&&&&\end{smallmatrix}\right)
&&&\pib_3=\left(\begin{smallmatrix}&&&&\,0&0&\,0&0\\&&&&\,0&0&-1&0\\&&&&\,0&0&\,0&0\\&&&&-1&0&\,0&0\\ \,0&\,0&\,0&\,0&&&&\\\,0&\,0&\,1&\,0&&&&\\\,0&\,0&\,0&\,0&&&&\\\,1&\,0&\,0&\,0&&&&\end{smallmatrix}\right)\quad.
\end{align*}
These matrices satisfy the relations
\begin{align*}
&\{\pi_i,\pi_j\}=0\,,\quad\{\pib_i,\pib_j\}=0,\\
&\{\pi_i,\pib_j\}=\begin{cases}0\quad  \text{if}\quad  i\neq j\\ 1\quad \text{if}\quad  i=j \end{cases}\quad.
\end{align*}
Using this we have 
\begin{multline*}
D\equiv X_1\pi_1+X_2 \pi_2+X_3 \pi_3 +X_1^{n-1}\,\pib_1+X_2^{n-1}\,\pib_2+X_3^{n-1}\,\pib_3
\\=
\left(\begin{smallmatrix}0&0&0&0&X_1&0&-X_2&X_3\\0&0&0&0&0&X_1&-X_3^{n-1}&-X_2^{n-1}\\0&0&0&0&X_2^{n-1}&X_3&X_1^{n-1}&0\\0&0&0&0&-X_3^{n-1}&X_2&0&X_1^{n-1}\\X_1^{n-1}&0&X_2&-X_3&0&0&0&0\\0&X_1^{n-1}&X_3^{n-1}&X_2^{n-1}&0&0&0&0\\-X_2^{n-1}&-X_3&X_1&0&0&0&0&0\\X_3^{n-1}&-X_2&0&X_1&0&0&0&0\end{smallmatrix}\right)
\end{multline*}
and
\begin{align*}
&\Psi_1\equiv\pi_1+X_1^{n-2}\pib_1=\left(\begin{smallmatrix}0&0&0&0&\phantom{(}1\,&\,0\,&\,0\,&\,0\,\\0&0&0&0&\,0&\phantom{(}1&\,0&\,0\\0&0&0&0&\,0&\,0&X_1^{n-2}&\,0\\0&0&0&0&\,0&\,0&\,0&X_1^{n-2}\\X_1^{n-2}&\,0\,&\,0\,&\,0\,&0&0&0&0\\\,0&X_1^{n-2}&\,0&\,0&0&0&0&0\\\,0&\,0&\phantom{(}1&\,0&0&0&0&0\\\,0&\,0&\,0&\phantom{(}1&0&0&0&0\end{smallmatrix}\right)\\\intertext{}
&\Psi_2\equiv\pi_2+X_2^{n-2}\pib_2=\left(\begin{smallmatrix}0&0&0&0&0&0&-1&0\\0&0&0&0&0&0&\,0&-X_2^{n-2}\\0&0&0&0&X_2^{n-2}&0&\,0&0\\0&0&0&0&0&\phantom{(}1&\,0&0\\ 0&\,0&\phantom{(}1&0&0&0&0&0\\0&\,0&0&X_2^{n-2}&0&0&0&0\\-X_2^{n-2}&\,0&0&0&0&0&0&0\\0&-1&0&0&0&0&0&0\end{smallmatrix}\right)\\\intertext{}
&\Psi_3\equiv\pi_3+X_3^{n-2}\pib_3=\left(\begin{smallmatrix}0&0&0&0&\,0&\,0&\,0&\,1\\0&0&0&0&\,0&\,0&-X_3^{n-2}&\,0\\0&0&0&0&\,0&\,1&\,0&\,0\\0&0&0&0&-X_3^{n-2}&\,0&\,0&\,0\\ 0&\,0&0&-1&0&0&0&\\0&\,0&X_3^{n-2}&\,0&0&0&0&0\\0&-1&0&\,0&0&0&0&0\\X_3^{n-2}&\,0&0&\,0&0&0&0&0\end{smallmatrix}\right)\qquad.\\
\end{align*}

\noindent Now we can obtain an explicit expression in matrix form for the boundary operators listed in Section \ref{s3ex}. The operators between the fractional branes $\mathcal{F}_\alpha$ can be rewritten as 
\begin{equation*}
\begin{split}
\intertext{}
&\Psi_1+\Psi_2+\Psi_3= \left(
\begin{smallmatrix}
0&0&0&0&1&0&-1&1\\
0&0&0&0&0&1&X_3^{n-2}&X_2^{n-2}\\
0&0&0&0&-X_2^{n-2}&1&-X_1^{n-2}&0\\
0&0&0&0&X_3^{n-2}&1&0&-X_1^{n-2}\\
-X_1^{n-2}&0&1&-1&0&0&0&0\\
0&-X_1^{n-2}&-X_3^{n-2}&-X_2^{n-2}&0&0&0&0\\
X_2^{n-2}&-1&1&0&0&0&0&0\\
-X_3^{n-2}&-1&0&1&0&0&0&0
\end{smallmatrix}\right)\\\intertext{}
&\Psi_1\Psi_2+\Psi_2\Psi_3+\Psi_3\Psi_1= \left(
\begin{smallmatrix}
0&1&1&1&0&0&0&0\\
-X_2^{n-2}X_3^{n-2}&0&X_3^{n-2}&-X_2^{n-2}&0&0&0&0\\
-X_1^{n-2}X_2^{n-2}&-X_1^{n-2}&0&X_2^{n-2}&0&0&0&0\\
-X_1^{n-2}X_3^{n-2}&X_1^{n-2}&-X_3^{n-2}&0&0&0&0&0\\
0&0&0&0&0&1&X_1^{n-2}&X_1^{n-2}\\
0&0&0&0&-X_2^{n-2}X_3^{n-2}&0&X_2^{n-2}X_3^{n-2}&-X_1^{n-2}X_2^{n-2}\\
0&0&0&0&-X_2^{n-2}&-1&0&X_2^{n-2}\\
0&0&0&0&-X_3^{n-2}&1&-X_3^{n-2}&0
\end{smallmatrix}\right)
\\\intertext{}
&\Psi_1\Psi_2\Psi_3= \left(
\begin{smallmatrix}
0&0&0&0&0&1&0&0\\
0&0&0&0&-X_2^{n-2}X_3^{n-2}&0&0&0\\
0&0&0&0&0&0&0&-X_1^{n-2}X_2^{n-2}\\
0&0&0&0&0&0&X_1^{n-2}X_3^{n-2}&0\\
0&-X_1^{n-2}&0&0&0&0&0&0\\
X_1^{n-2}X_2^{n-2}X_3^{n-2}&0&0&0&0&0&0&0\\
0&0&0&X_2^{n-2}&0&0&0&0\\
0&0&-X_3^{n-2}&0&0&0&0&0
\end{smallmatrix}\right)\intertext{}
\end{split}
\end{equation*}

\noindent For the operators of  $\mathcal{F}_{\kern-.10em\begin{picture}(6,8)
\put(0,3){\line(1,0){6}}\put(0,0){\line(1,0){3}}\put(0,6){\line(1,0){6}}\put(0,0){\line(0,1){6}}\put(3,0){\line(0,1){6}}\put(6,3){\line(0,1){3}}
\end{picture}}$ the matrices become quite large, so we will use the notation 
$$\mathcal{O}=\left(\begin{smallmatrix}\phi_-&\psi_-\\\psi_+&\phi_+\end{smallmatrix}\right)$$ for a generic boundary operator $\mathcal{O}$ and as usual $\phi$ denotes the even part and $\psi$ the odd part. We can then rewrite the operators in \eqref{sol1} as:
\begin{align*}\\
\text{i.\quad }\quad&\phi=(\pi_1\pib_1+\pib_1\pi_1)(\pi_2\pib_2+\pib_2\pi_2)(\pi_3\pib_3+\pib_3\pi_3)\equiv\II_{16}\intertext{}\\
\text{ii.\quad }\quad&\phi_-\equiv
\left(
\begin{smallmatrix}
0&0&0&0&0&1&1&1\\
0&0&0&0&-X_2^{n-2}X_3^{n-2}&0&X_3^{n-2}&-X_2^{n-2}\\
0&0&0&0&-X_1^{n-2}X_2^{n-2}&-X_1^{n-2}&0&X_2^{n-2}\\
0&0&0&0&-X_1^{n-2}X_3^{n-2}&X_1^{n-2}&-X_3^{n-2}\\
0&-1&-1&-1&0&0&0&0\\
X_2^{n-2}X_3^{n-2}&0&-X_3^{n-2}&X_2^{n-2}&0&0&0&0\\
X_1^{n-2}X_2^{n-2}&X_1^{n-2}&0&-X_2^{n-2}&0&0&0&0\\
X_1^{n-2}X_3^{n-2}&-X_1^{n-2}&X_3^{n-2}&0&0&0&0&0\\
\end{smallmatrix}\right)\\
&\phi_+\equiv
\left(
\begin{smallmatrix}
0&0&0&0&0&1&X_1^{n-2}&X_1^{n-2}\\
0&0&0&0&-X_2^{n-2}X_3^{n-2}&0&X_2^{n-2}X_3^{n-2}&-X_1^{n-2}X_2^{n-2}\\
0&0&0&0&-X_2^{n-2}&-1&0&X_2^{n-2}\\
0&0&0&0&-X_3^{n-2}&1&-X_3^{n-2}&0\\
0&-1&-X_1^{n-2}&-X_1^{n-2}&0&0&0&0\\
X_2^{n-2}X_3^{n-2}&0&-X_2^{n-2}X_3^{n-2}&X_1^{n-2}X_2^{n-2}&0&0&0&0\\
X_2^{n-2}&1&0&-X_2^{n-2}&0&0&0&0\\
X_3^{n-2}&-1&X_3^{n-2}&0&0&0&0&0
\end{smallmatrix}\right)\\\intertext{}
\text{iii.\quad }\quad&\phi_-\equiv
\left(
\begin{smallmatrix}
0&\sqrt{3}&0&-\sqrt{3}&0&0&3&0\\
-\sqrt{3}X_2^{n-2}X_3^{n-2}&0&-\sqrt{3}X_3^{n-2}&0&0&0&0&-3X_2^{n-2}\\
0&\sqrt{3}X_1^{n-2}&0&\sqrt{3}X_2^{n-2}&-3X_1^{n-2}X_2^{n-2}&0&0&0\\
\sqrt{3}X_1^{n-2}X_3^{n-2}&0&-\sqrt{3}X_3^{n-2}&0&0&3X_1^{n-2}&0&0\\
0&-2&1&-2&0&-\sqrt{3}&0&\sqrt{3}\\
2X_2^{n-2}X_3^{n-2}&0&-2X_3^{n-2}&-X_2^{n-2}&\sqrt{3}X_2^{n-2}X_3^{n-2}&0&\sqrt{3}X_3^{n-2}&0\\
-X_1^{n-2}X_2^{n-2}&2X_1^{n-2}&0&-2X_2^{n-2}&0&-\sqrt{3}X_1^{n-2}&0&-\sqrt{3}X_2^{n-2}\\
2X_1^{n-2}X_3^{n-2}&X_1^{n-2}&2X_3^{n-2}&0&-\sqrt{3}X_1^{n-2}X_3^{n-2}&0&\sqrt{3}X_3^{n-2}&0
\end{smallmatrix}\right)\\
&\phi_+\equiv
\left(
\begin{smallmatrix}
0&\sqrt{3}&0&-\sqrt{3}X_1^{n-2}&0&0&3X_1^{n-2}&0\\
-\sqrt{3}X_2^{n-2}X_3^{n-2}&0&-\sqrt{3} X_1^{n-2}X_3^{n-2}&0&0&0&0&-3X_1^{n-2}X_2^{n-2}\\
0&\sqrt{3}&0&\sqrt{3}X_2^{n-2}&-3X_2^{n-2}&0&0&0\\
\sqrt{3}X_3^{n-2}&0&-\sqrt{3}X_3^{n-2}&0&0&3&0&0\\
0&-2&X_1^{n-2}&-2X_1^{n-2}&0&-\sqrt{3}&0&\sqrt{3}X_1^{n-2}\\
2X_2^{n-2}X_3^{n-2}&0&-2X_1^{n-2}X_3^{n-2}&-X_1^{n-2}X_2^{n-2}&\sqrt{3}X_2^{n-2}X_3^{n-2}&0&\sqrt{3}X_1^{n-2}X_3^{n-2}&0\\
-X_2^{n-2}&2&0&-2X_2^{n-2}&0&-\sqrt{3}&0&-\sqrt{3}X_2^{n-2}\\
2X_3^{n-2}&1&2X_3^{n-2}&0&-\sqrt{3}X_3^{n-2}&0&\sqrt{3}X_3^{n-2}&0
\end{smallmatrix}\right)
\\\intertext{}
\text{iv.\quad }\quad&\psi_-\equiv
\left(
\begin{smallmatrix}
1&0&-1&1&0&0&0&0\\
0&1&X_3^{n-2}&X_2^{n-2}&0&0&0&0\\
-X_2^{n-2}&1&-X_1^{n-2}&0&0&0&0&0\\
X_3^{n-2}&1&0&-X_1^{n-2}&0&0&0&0\\
0&0&0&0&1&0&-1&1\\
0&0&0&0&0&1&X_3^{n-2}&X_2^{n-2}\\
0&0&0&0&-X_2^{n-2}&1&-X_1^{n-2}&0\\
0&0&0&0&X_3^{n-2}&1&0&-X_1^{n-2}\\
\end{smallmatrix}\right)\\
&\psi_+\equiv
\left(
\begin{smallmatrix}
-X_1^{n-2}&0&1&-1&0&0&0&0\\
0&-X_1^{n-2}&-X_3^{n-2}&-X_2^{n-2}&0&0&0&0\\
X_2^{n-2}&-1&1&0&0&0&0&0\\
-X_3^{n-2}&-1&0&1&0&0&0&0\\
0&0&0&0&-X_1^{n-2}&0&1&-1\\
0&0&0&0&0&-X_1^{n-2}&-X_3^{n-2}&-X_2^{n-2}\\
0&0&0&0&X_2^{n-2}&-1&1&0\\
0&0&0&0&-X_3^{n-2}&-1&0&1
\end{smallmatrix}\right)
\\\intertext{}
\text{v.\quad }\quad&\psi_-\equiv
\left(
\begin{smallmatrix}
0&0&0&0&0&1&0&0\\
0&0&0&0&-X_2^{n-2}X_3^{n-2}&0&0&0\\
0&0&0&0&0&0&0&-X_1^{n-2}X_2^{n-2}\\
0&0&0&0&0&0&X_1^{n-2}X_3^{n-2}&0\\
0&-1&0&0&0&0&0&0\\
X_2^{n-2}X_3^{n-2}&0&0&0&0&0&0&0\\
0&0&0&X_1^{n-2}X_2^{n-2}&0&0&0&0\\
0&0&-X_1^{n-2}X_3^{n-2}&0&0&0&0&0\\
\end{smallmatrix}\right)\\
&\psi_+\equiv
\left(
\begin{smallmatrix}
0&0&0&0&0&-X_1^{n-2}&0&0\\
0&0&0&0&X_1^{n-2}X_2^{n-2}X_3^{n-2}&0&0&0\\
0&0&0&0&0&0&0&X_2^{n-2}\\
0&0&0&0&0&0&-X_3^{n-2}&0\\
0&X_1^{n-2}&0&0&0&0&0&0\\
-X_1^{n-2}X_2^{n-2}X_3^{n-2}&0&0&0&0&0&0&0\\
0&0&0&-X_2^{n-2}&0&0&0&0\\
0&0&X_3^{n-2}&0&0&0&0&0
\end{smallmatrix}\right)\\\\
\intertext{Finally, the operators between $\mathcal{F}_0$ and $\mathcal{F}_{\kern-.10em\begin{picture}(6,8)
\put(0,3){\line(1,0){6}}\put(0,0){\line(1,0){3}}\put(0,6){\line(1,0){6}}\put(0,0){\line(0,1){6}}\put(3,0){\line(0,1){6}}\put(6,3){\line(0,1){3}}
\end{picture}}$  in \eqref{sol2} take the form:}
\\
\text{i.\quad }\quad&\psi=\left(\begin{smallmatrix}
0&0&0&0&1&0&-1&-2\\
0&0&0&0&0&1&-2X_3^{n-2}&X_2^{n-2}\\
0&0&0&0&-2X_2^{n-2}&-2&-X_1^{n-2}&0\\
0&0&0&0&-2X_3^{n-2}&1&0&-X_1^{n-2}\\
0&0&0&0&-\sqrt{3}&0&-\sqrt{3}&0\\
0&0&0&0&0&-\sqrt{3}&0&\sqrt{3}X_2^{n-2}\\
0&0&0&0&-\sqrt{3}X_2^{n-2}&0&\sqrt{3}X_1^{n-2}&0\\
0&0&0&0&0&\sqrt{3}&0&\sqrt{3}X_1^{n-2}\\
-X_1^{n-2}&0&1&2&0&0&0&0\\
0&-X_1^{n-2}&2X_3^{n-2}&-X_2^{n-2}&0&0&0&0\\
X_2^{n-2}&2&1&0&0&0&0&0\\
2X_3^{n-2}&-1&0&1&0&0&0&0\\
\sqrt{3}X_1^{n-2}&0&\sqrt{3}&0&0&0&0&0\\
0&\sqrt{3}X_1^{n-2}&0&-\sqrt{3}X_2^{n-2}&0&0&0&0\\
\sqrt{3}X_2^{n-2}&0&-\sqrt{3}&0&0&0&0&0\\
0&-\sqrt{3}&0&-\sqrt{3}&0&0&0&0
\end{smallmatrix}\right)\\\\
\text{ii.\quad }\quad&\phi=\left(
\begin{smallmatrix}
0&-1&2&-1&0&0&0&0\\
X_2^{n-2}X_3^{n-2}&0&-X_3^{n-2}&-2X_2^{n-2}&0&0&0&0\\
-2X_1^{n-2}X_2^{n-2}&X_3^{n-2}&0&-X_2^{n-2}&0&0&0&0\\
X_1^{n-2}X_3^{n-2}&2X_1^{n-2}&0&-X_2^{n-2}&0&0&0&0\\
X_1^{n-2}X_3^{n-2}&2X_1^{n-2}&X_3^{n-2}&0&0&0&0&0\\
0&\sqrt{3}&0&\sqrt{3}&0&0&0&0\\
-\sqrt{3}X_2^{n-2}X_3^{n-2}&0&-\sqrt{3}X_3^{n-2}&0&0&0&0&0\\
0&\sqrt{3}X_1^{n-2}&0&\sqrt{3}X_2^{n-2}&0&0&0&0\\
\sqrt{3}X_1^{n-2}X_3^{n-2}&0&-\sqrt{3}X_3^{n-2}&0&0&0&0&0\\
0&0&0&0&0&-1&2X_1^{n-2}&-X_1^{n-2}\\
0&0&0&0&X_2^{n-2}X_3^{n-2}&0&-X_1^{n-2}X_3^{n-2}-2X_1^{n-2}X_2^{n-2}\\
0&0&0&0&-2X_2^{n-2}&1&0&-X_2^{n-2}\\
0&0&0&0&X_3^{n-2}&2&X_3^{n-2}&0\\
0&0&0&0&0&\sqrt{3}70&-\sqrt{3}X_1^{n-2}\\
0&0&0&0&-\sqrt{3}X_2^{n-2}X_3^{n-2}&0&-\sqrt{3}X_1^{n-2}X_3^{n-2}&0\\
0&0&0&0&0&\sqrt{3}&0&\sqrt{3}X_2^{n-2}\\
0&0&0&0&\sqrt{3}X_3^{n-2}&0&-\sqrt{3}X_3^{n-2}&0\\
\end{smallmatrix}
\right)\quad.
\end{align*}

\end{appendix}

\end{document}